\documentclass[10pt]{article}
\usepackage{geometry}
\usepackage{fancyhdr}
\usepackage{amssymb}
\usepackage{graphicx}
\usepackage{tabu}
\usepackage[caption=false,font=footnotesize]{subfig}

\usepackage{url}
\urldef{\mailsa}\path|auyar@iu.edu|

\begin{document}


\title{Intel Optane DCPMM and Serverless Computing}

%
%
\author{
\resizebox{0.55\linewidth}{!}{
\begin{tabular}{cccc}
Ahmet Uyar$^1$ & Selahattin Akkas$^1$ &Jiayu Li$^1$ & Judy Fox$^2$\\
\multicolumn{2}{c}{$^{1}$Indiana University} & \multicolumn{2}{c}{$^{2}$University of Virginia} \\
\multicolumn{4}{c}{$^1$\{auyar,sakkas,jl145\}@iu.edu}\\ \multicolumn{4}{c}{$^2$\{ckw9mp\}@virginia.edu}
\end{tabular}}\\
}

\maketitle

\begin{abstract}
This report describes 1) how we use Intel’s Optane DCPMM in the memory Mode. We investigate the the scalability of applications on a single Optane machine, using Subgraph counting as memory-intensive graph problem. We test with various input graph and subtemplate sizes to determine its performance for different memory and CPU loads, as well as a comparison of performance on a single node Optane with a distributed set of nodes in a cluster using MPI. 2) We investigate the end-to-end execution delays in serverless computing and study concurrent function executions with cold starts. In future work, we will show that persistent memory machines may significantly improve concurrent function invocations in serverless computing including Amazon Lambda, Microsoft Azure Functions, Google Cloud Functions and IBM Cloud Functions (Apache OpenWhisk).

\end{abstract}

\section{Introduction}

Subgraph counting is an NP-hard and memory-intensive graph problem. It is the process of counting a given subgraph T in a large graph G. It has many use cases in diverse domains from social network analysis \cite{sna} to bioinformatics \cite{bioinfo}. Since it is an NP-hard problem, several approximation algorithms have been developed. We use the color-coding approximation algorithm \cite{cc}. In addition, several versions of this problem have been studied in the literature. We use the version \cite{countingGraphlets} that counts tree templates in graphs. 
\begin{figure}[ht]
     \centering
     \includegraphics[width=0.6\linewidth]{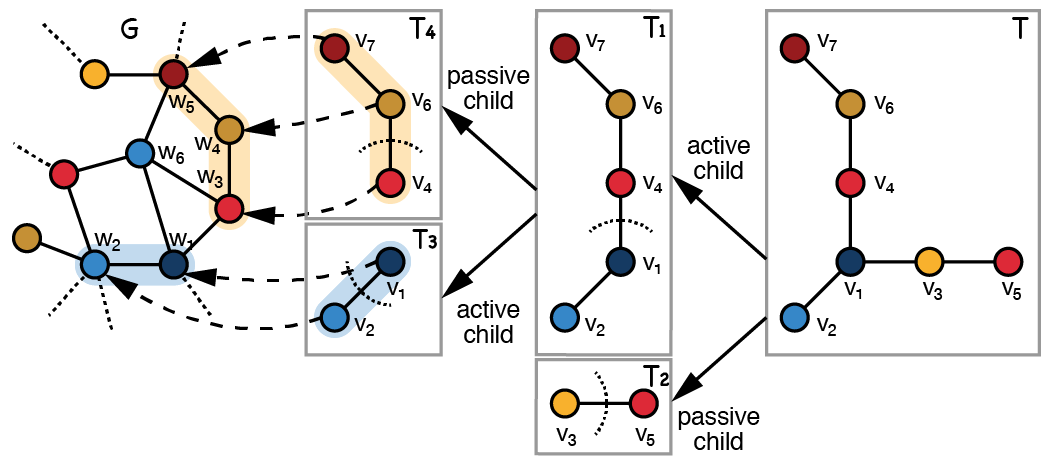}
     \caption{Illustration of the template partitioning within a
     colored input $G=(V,E)$}
     \label{fig:partitionTemplates}
 \end{figure}
Intel\textregistered\ Optane\textsuperscript{TM} DC Persistent Memory Module is a nonvolatile storage module supporting byte-granularity accesses and persistency \cite{optanePerf}. It is a module that goes into a tier in-between SSD and DRAM. It is used in either of two modes: Memory mode or App-direct mode. In memory-mode, persistent memory module is used as the main memory of the machine. It provides a much larger capacity than DRAM with slightly lower performance. In this mode, DRAM functions as the fourth level cache and no persistency provided for the data in memory. In app-direct mode, persistent memory module functions as the byte-addressable persistent storage device, and it provides much faster and higher bandwidth data accesses than SSDs.

The experiments are conducted on two Optane machines.  They have 2x48 core Intel\textregistered\ Xeon\textregistered\ Platinum 8260L CPU @ 2.40GHz processors with 384GB of DRAM and 4TB of Persistent Memory. Since the subgraph counting algorithm is memory-intensive, we use those machines in the memory mode. They provide 4TB of main memory and 384GB of the fourth level cache. We investigate the scaling efficiency of the subgraph counting algorithm on a single Optane machine with an increasing number of OpenMP threads. We also compare the performance of the multi-threaded version of the subgraph counting algorithm on a single Optane machine to the distributed version of the same algorithm on a cluster. 
\begin{table}[h!]
\centering
\begin{tabular}{llllll} 
 \hline
Type & CPU  &Frequency  & Cores & DRAM & Persistent Memory \\ 
Optane & Xeon Platinum 8260L ~& 2.4GHz&24  & 384GB ~ & 4TB \\
Haswell & E5-2670 v3 & 2.3GHz &12 & 128GB & \\
 \hline
\end{tabular}
\caption{Hardware}
\label{table:hardware}
\end{table}

Serverless Computing is an emerging paradigm for cloud applications \cite{sc1}, \cite{sc2}. It promises a simplified programming model and maintenance-free deployment. Developers only write functions and deploy them on the cloud. Cloud providers take care of resource provisioning, monitoring, maintenance, scalability, and fault-tolerance. Currently, all major cloud vendors provide serverless cloud infrastructure. Amazon provides AWS Lambda\footnote{https://aws.amazon.com/lambda/}, Google provides Google Cloud Functions\footnote{https://cloud.google.com/functions/}, Microsoft provides Azure Functions\footnote{https://azure.microsoft.com/en-us/services/functions} and IBM provides IBM Cloud Functions\footnote{https://www.ibm.com/cloud/functions}. IBM open-sourced its serverless platform as OpenWhisk\footnote{https://openwhisk.apache.org/} and Microsoft open-sourced its Azure Functions\footnote{https://github.com/Azure/Azure-Functions}. 

Serverless computing was the fastest-growing cloud service for the second year in a row, according to Rightscale 2019 State Of The Cloud Report From Flexera \cite{cloudstate}. It has a wide range of use cases \cite{sless-wp}. Some of its recommended use cases can be listed as the following: multimedia processing, HTTP REST APIs and web applications, event-triggered processing in databases, performing analytics on IoT sensor inputs, handling stream processing, managing single time extract-transform-load jobs, providing cognitive computing via chatbots, serving machine learning and AI models, etc. 

Serverless computing platforms employ containers to execute functions. A separate container is constructed for each function. While it is very fast to execute functions when there is a ready container, it is much slower to cold start containers and executes the functions. We conducted tests to show the end-to-end function execution delays and concurrent function execution delays in OpenWhisk serverless computing system. We conducted the tests using both HDD and SSD drives. We demonstrate the impact of faster storage on these services. Although we have not performed these tests on the Optane machine, we make the case that serverless computing systems would benefit significantly by using faster and higher bandwidth persistent memory machines. 

The report is organized as follows: we present the performance tests results related to the subgraph counting algorithm on Optane machine in section~\ref{subCount}. We evaluate the performance of serverless computing in section~\ref{serverless}. Conclusions and future work are at section~\ref{conclusions}. 

\section{Performance Tests for Subgraph Counting} \label{subCount}

A parallel version of the subgraph counting algorithm has been developed at Digital Science Center at Indiana University using OpenMP threads and AVX2 vector instructions for multi-core machines. A distributed version is also developed using MPI to count larger subgraphs in clusters \cite{SubGraph2Vec}. We are using these implementations to run the subgraph counting on single and distributed nodes. 

To conduct the performance tests, we use the graph datasets shown at Table ~\ref{table:datasets}. First graph dataset (nyc.graph) is a real social network graph data with 17M vertices and 480M edges from \cite{nyc}. Other two datasets are synthetic datasets generated by RMAT graph generation algorithm \cite{rmat} using PaRMAT tool \footnote{https://github.com/farkhor/PaRMAT} with a larger number of vertices and edges. We count subgraph tree templates that have vertices ranging from 3 to 16. 

\begin{table}[h!]
\centering
\begin{tabu} to 0.8\textwidth {|X[l] |X[l] | X[l]|} 
 \hline
 Dataset & Vertices & Edges \\ 
 \hline
 nyc.graph & 17,876,290 & 480,093,705 \\ 
 \hline
 RMAT1 (k=3) & 30M & 800M  \\
 \hline
 RMAT2 (k=1) & 40M & 1000M \\
 \hline
\end{tabu}
\caption{Graph datasets used in tests}
\label{table:datasets}
\end{table}

\subsection{Scaling with OpenMP Threads} \label{scalingWithOpenMP}

We first test the scalability of the subgraph counting algorithm on a single Optane machine with an increasing number of OpenMP threads. We conduct the subgraph counting tests on nyc.graph dataset. We count two different input subgraphs with different peak memory requirements. First input subgraph has 7 vertices. Counting this subgraph requires 28GB of peak memory. Therefore, all data can fit into the DRAM of the machine easily. There is no need to move the data between DRAM and persistent memory during the computation. Second subgraph has 15 vertices. Counting that subgraph requires 742GB of peak memory. This peak memory is much more than the DRAM capacity of the machine that is 384GB. Therefore, we expect a lot of DRAM to persistent memory data movement during the computation. We want to determine the scalability of the subgraph counting algorithm with both types of memory loads. 

\begin{figure*}[htb]
\centering
 \includegraphics[width=0.4\linewidth]{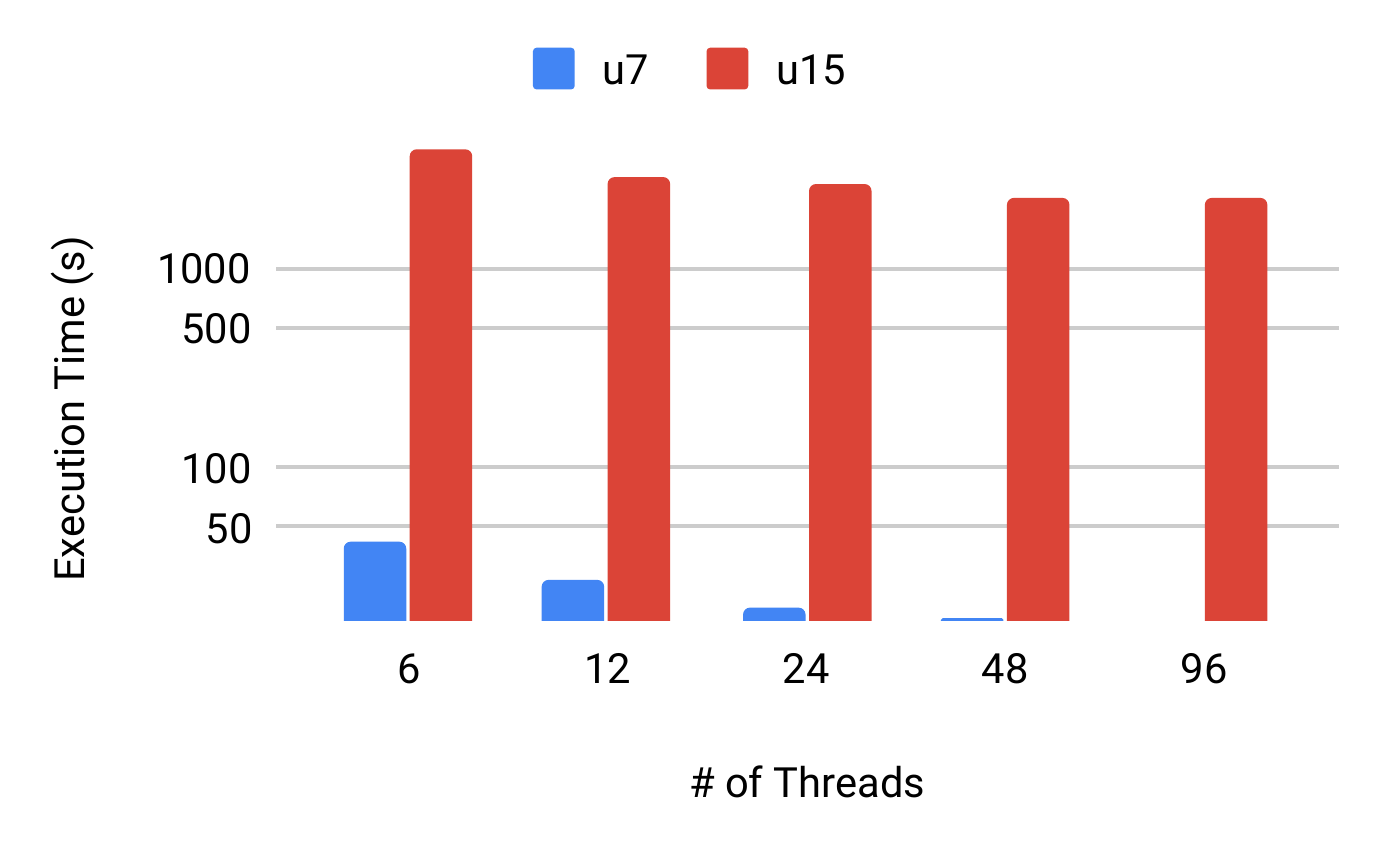}
\caption{Scaling subgraph counting with OpenMP threads on a single Optane machine}
\label{fig:openMPScaling}
\end{figure*}

Figure~\ref{fig:openMPScaling} shows the results of the scalability tests. In the first task, we started with 1 thread and increased the number of threads up to 96. In the second task, we started with 6 threads and increased it up to 96. The reason for starting from 6 threads in the second task is twofold. First, our focus is on the scalability of this algorithm with a higher number of threads. We focus on speeding up this algorithm when all the resources of the machine are fully utilized. Second, running this algorithm on 6 threads already takes 3961 seconds (66min). Running it on a single thread would have taken around 5 hours. 

Both results show that the algorithm runs faster as the number of OpenMP threads increases. However, the rate of speedup decreases as the number of threads increases. Since the Optane machine has 48 physical cores, the algorithm does not run any faster for 96 threads. Executions times are very similar for 48 and 96 threads in both tasks. 

\begin{figure}[b]
\centering
\includegraphics[width=0.4\textwidth]{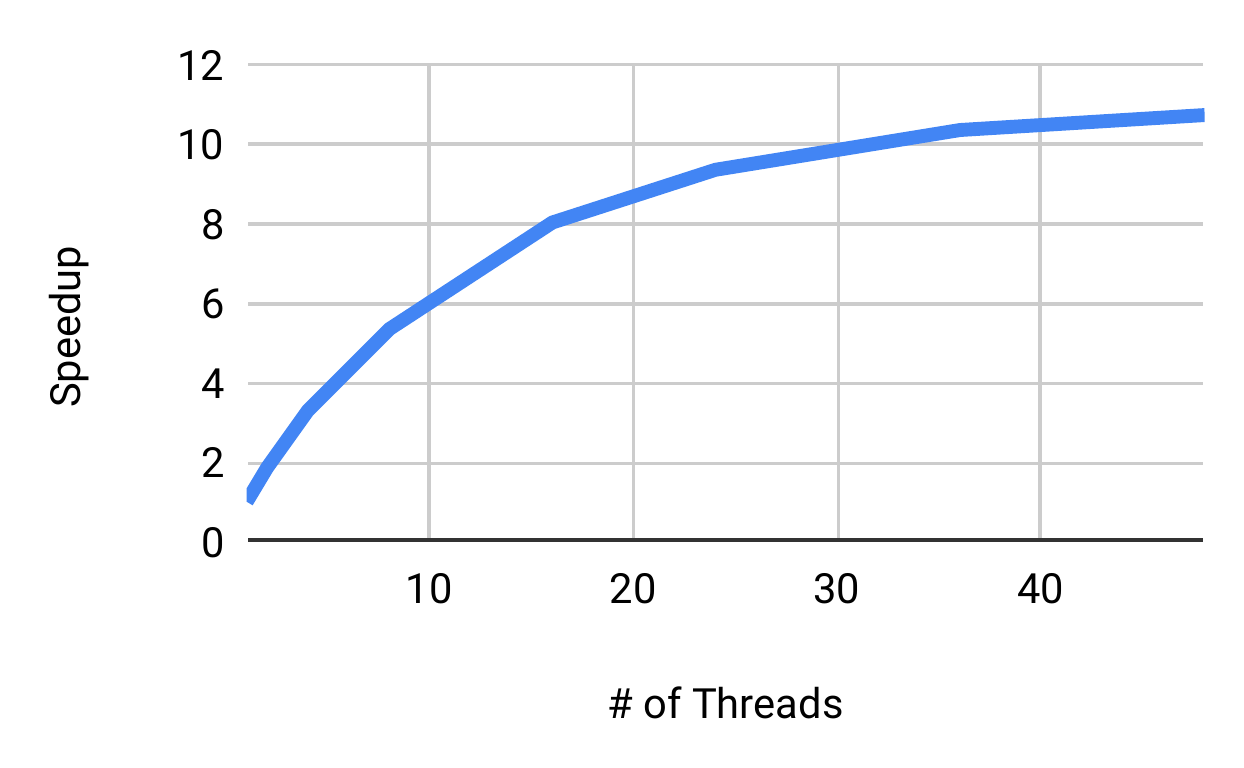} 
\caption{Speedup at a single Optane machine while counting 7 vertices subgraph in nyc.graph.}
\label{fig:threadScalingEfficiency}
\end{figure}

Figure~\ref{fig:threadScalingEfficiency} shows the efficiency of multi-threading for the 7 vertices input subgraph. We divide the speedup from single-threaded version to multi-threaded version by the number of threads. It shows the efficiency of parallel cores with respect to the single-threaded version. The value of 1 shows that all used cores are utilized as good as the single-threaded version. Multi-threading efficiency decreases rapidly as the number of threads increases. The efficiency is at 0.5 for 16 threads. It is at 0.22 for 48 threads. Although the decrease in efficiency is expected as the number of threads increases, we believe that better efficiency values should be attainable.  

\begin{figure*}[!tb]
\centering
\subfloat[Average DRAM Bandwidth]{\includegraphics[width=0.33\linewidth]{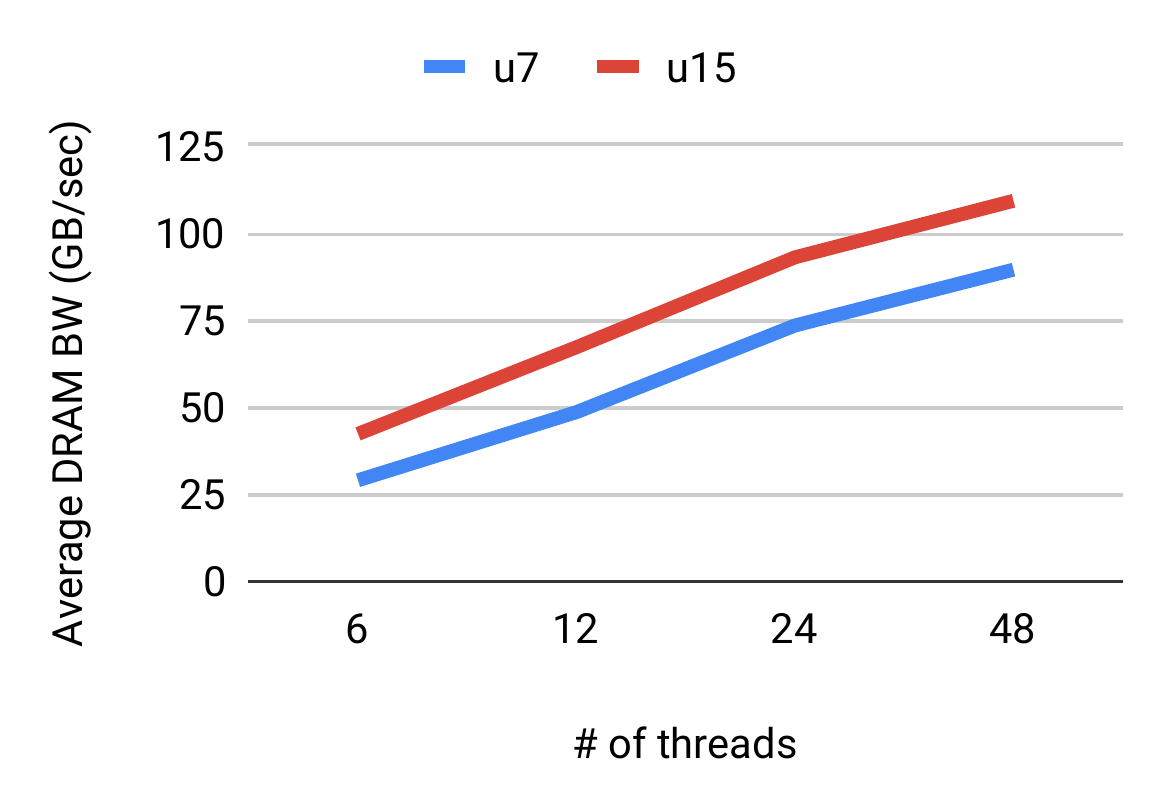}
\label{subfig:openMP-vtune-a}}
\subfloat[High DRAM Bandwith]{\includegraphics[width=0.33\linewidth]{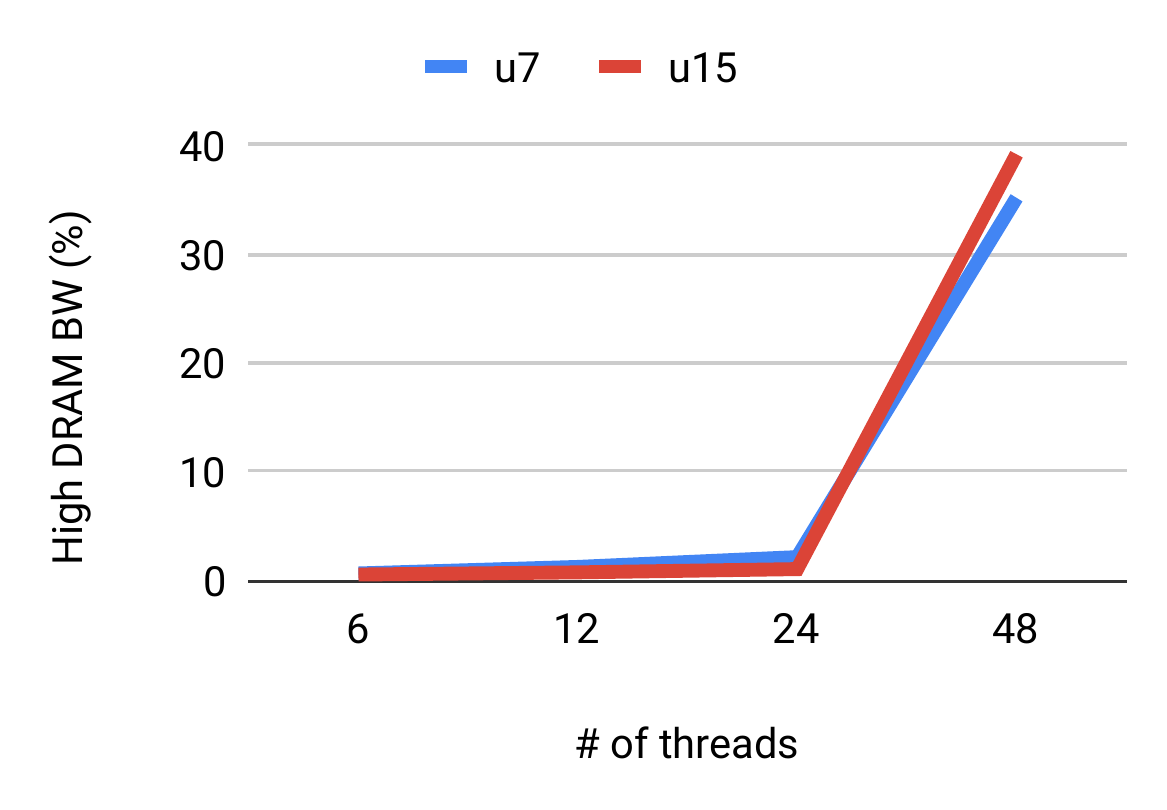}
\label{subfig:openMP-vtune-b}}
\vfill
\subfloat[DRAM Bound Clockticks]{\includegraphics[width=0.33\linewidth]{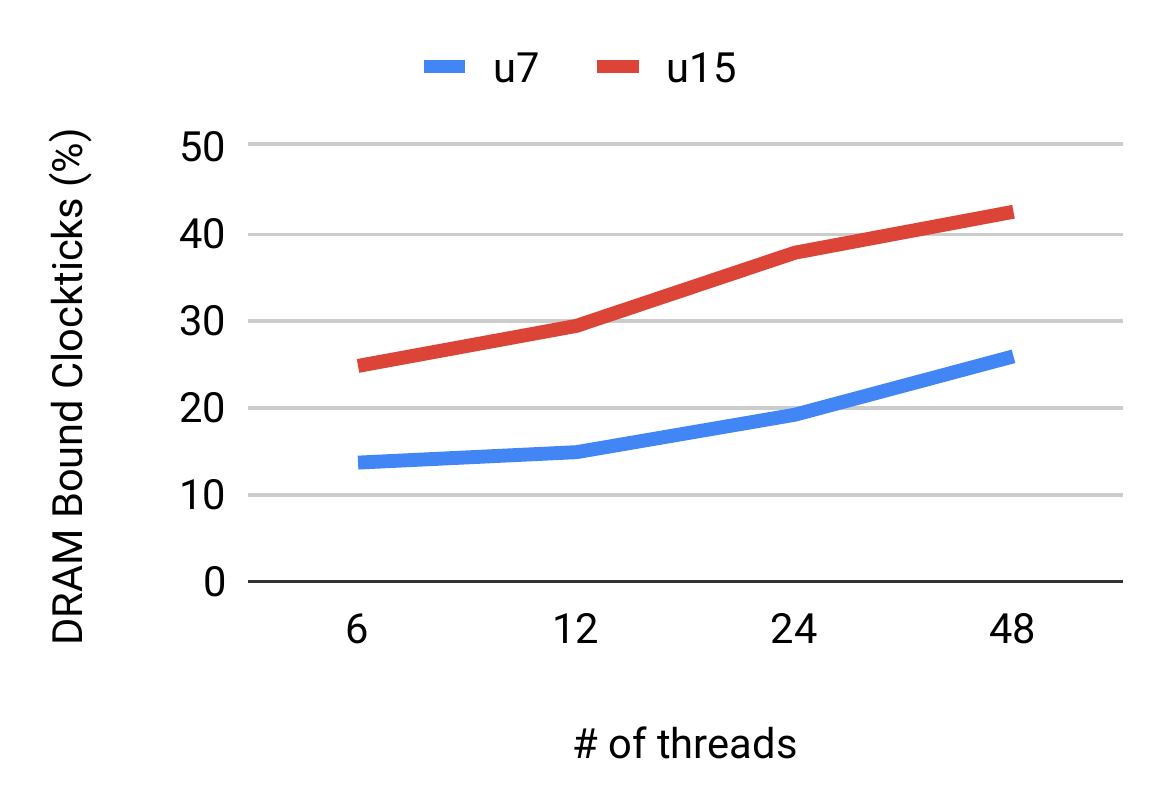}
\label{subfig:openMP-vtune-c}}
\subfloat[NUMA Remote Accesses]{\includegraphics[width=0.33\linewidth]{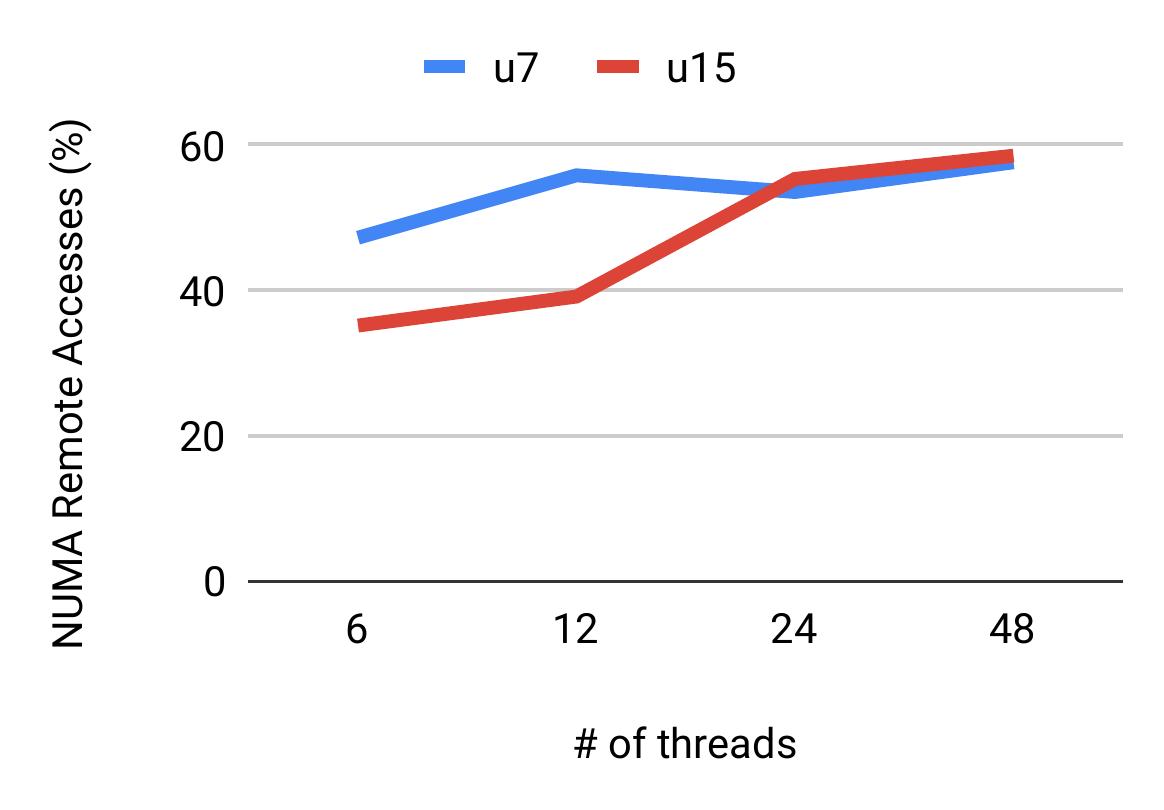}
\label{subfig:openMP-vtune-d}}
\caption{Performance metrics of Optane when counting 7 and 15 vertices subgraphs}
\label{fig:openMP-vtune}
\end{figure*}

We examined various hardware performance metrics of the Optane machine using Intel Vtune Amplifier for both subgraph counting tasks. The results are shown in figure~\ref{fig:openMP-vtune}. In general, this is a memory-intensive problem, and these hardware metrics demonstrate that.

Figure~\ref{subfig:openMP-vtune-a} shows average DRAM bandwidth for both tasks. As the number of threads increases, the average DRAM bandwidth also increases. Particularly after 24 threads, the machine operates in a high bandwidth range for a significant percentage of time as shown in Figure~\ref{subfig:openMP-vtune-b}. Figure~\ref{subfig:openMP-vtune-c} shows percentage of DRAM bound clockticks. Both tasks have very high DRAM bound clockticks. These values show that CPUs often stall for DRAM accesses. 

This Optane machine has two CPU sockets. It has NUMA memory architecture. Both sockets have their own DRAMs. When a thread on the first socket tries to access the data on the DRAM of the second socket, it incurs additional delays. It is called a NUMA remote access. Figure~\ref{subfig:openMP-vtune-d} shows the percentages of NUMA remote accesses. In both tasks, this percentage is very high. They are around 50\%. This means that executing threads are waiting longer for half of the memory accesses. On the contrary to other metrics, this metric does not have a clear pattern based on the number of threads. It goes up and down. This is the result of OpenMP thread assignments to cores in two sockets by the operating system. This assignment should be optimized to improve the performance of this algorithm in this machine. 

To improve the scalability of this problem, two types of improvements needed: a) locality of data in DRAM should be improved and consequently the percentages of DRAM bound instructions shall be reduced. b) Percentages of NUMA remote accesses should be reduced by assigning threads to cores more efficiently. 

\subsection{Scaling with MPI Processes} \label{scalingWithMPI}

In this section, we test whether running distributed version of the subgraph counting algorithm with multiple MPI processes on a single Optane machine improves memory utilization and provides faster executions. We use Intel MPI library to perform the tests. 

\begin{figure}[b]
\centering
\includegraphics[width=0.4\textwidth]{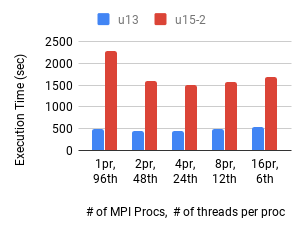} 
\caption{Scaling with MPI processes at an Optane machine}
\label{fig:mpiScaling}
\end{figure}

We count the subgraphs with 13 and 15 vertices in nyc.graph. The first task uses 254GB of peak memory, and the second task uses 745GB of peak memory in distributed mode. While all the data of the first task fits into the DRAM of the machine, the data of the second one spills over to the persistent memory. In each run, the number of MPI processes x the number of OpenMP threads per MPI process equals 96. So, we would like to utilize all available cores on the machine. Figure~\ref{fig:mpiScaling} shows the test results.

First bars show the result from non-distributed version with one process and 96 threads. In this case, no MPI is involved. They are the reference execution times for comparing to distributed versions. Other bars show the results from distributed version with 2 or more MPI processes on the same machine. These results indicate that using multiple MPI processes on an Optane machine usually improves the execution times. When four MPI processes are used with 24 threads on each, the algorithm performs the best in both tasks. Compared to the single process case, 4 MPI processes run 9\% faster for 13 vertices subgraph and 34\% faster for 15 vertices subgraph. It seems that the efficiency of the single process subgraph counting algorithm decreases as the memory usage rate increases. The primary reason for the better performance of the distributed version is the better memory utilization, as we will see in the figures of performance metrics below. 

We should also point out the fact that the multi-process distributed version introduces some inter-process communication and synchronization overheads. This is evident in counting the subgraph with 13 vertices. As the number of MPI processes increases to 8, inter-process communication and synchronization overheads matches the gains achieved by the better memory utilization and both single process version and the distributed version run with similar execution times. For 16 MPI processes, the distributed version provides slower performance. However, when counting the subgraph with 15 vertices, even when using 16 MPI processes, distributed version runs faster than the single process version. In this case, the efficiency gained by better memory utilization surpasses the costs of inter-process communication and synchronization overheads. 

\begin{figure*}[bt]
\centering
\subfloat[Average DRAM Bandwidth]{\includegraphics[width=0.33\linewidth]{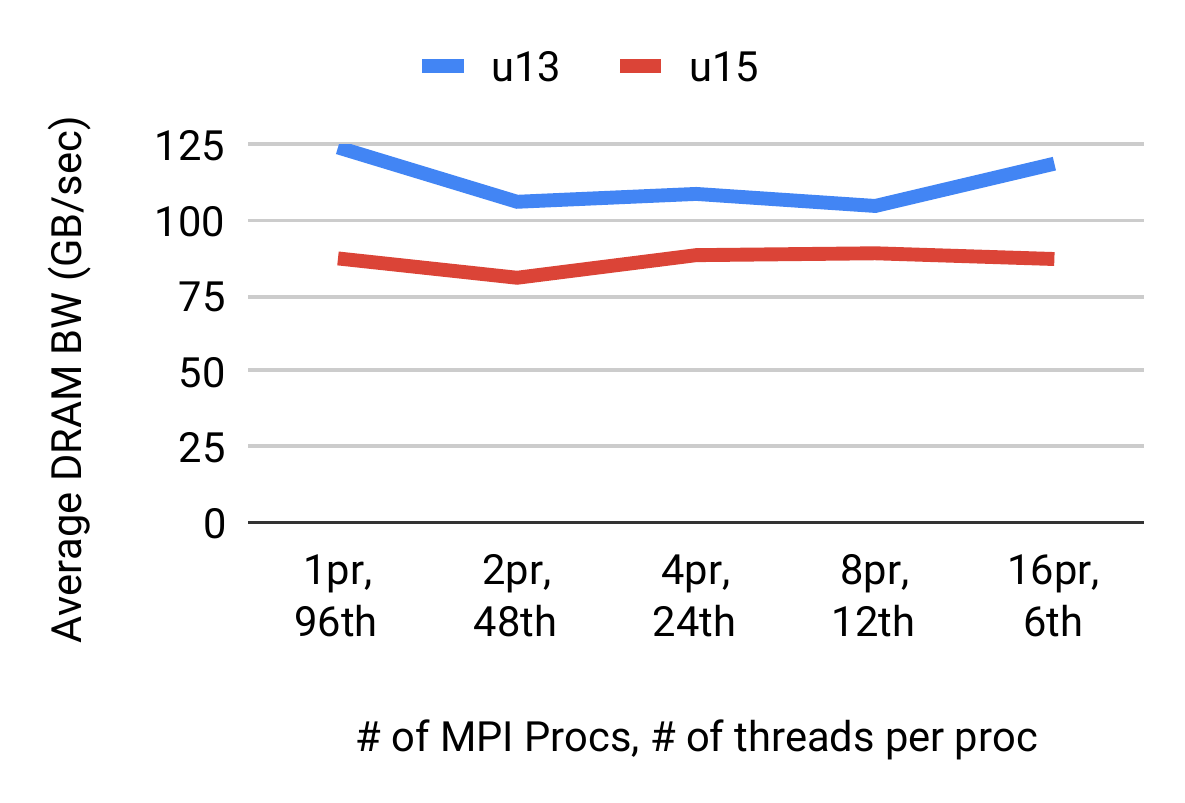}
\label{subfig:mpi-vtune-a}}
\subfloat[High DRAM Bandwidth]{\includegraphics[width=0.33\linewidth]{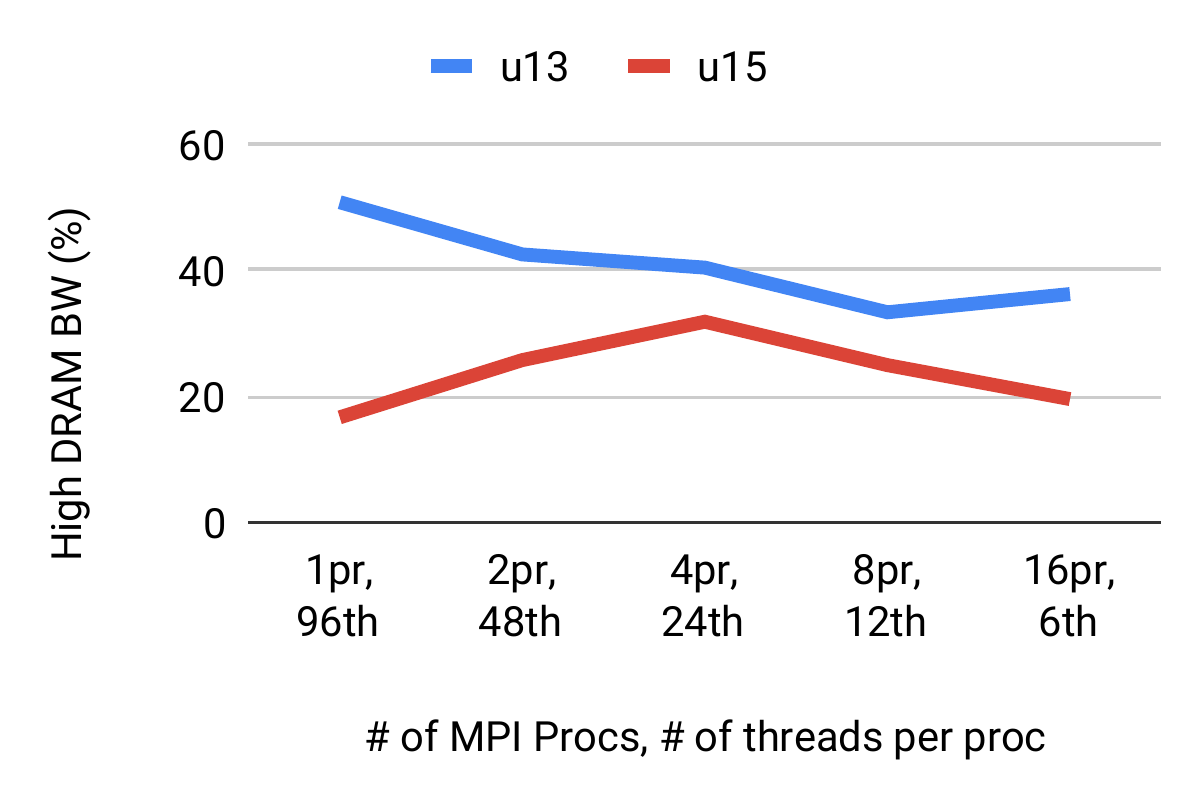}
\label{subfig:mpi-vtune-b}}
\vfill
\subfloat[DRAM Bound Clockticks]{\includegraphics[width=0.33\linewidth]{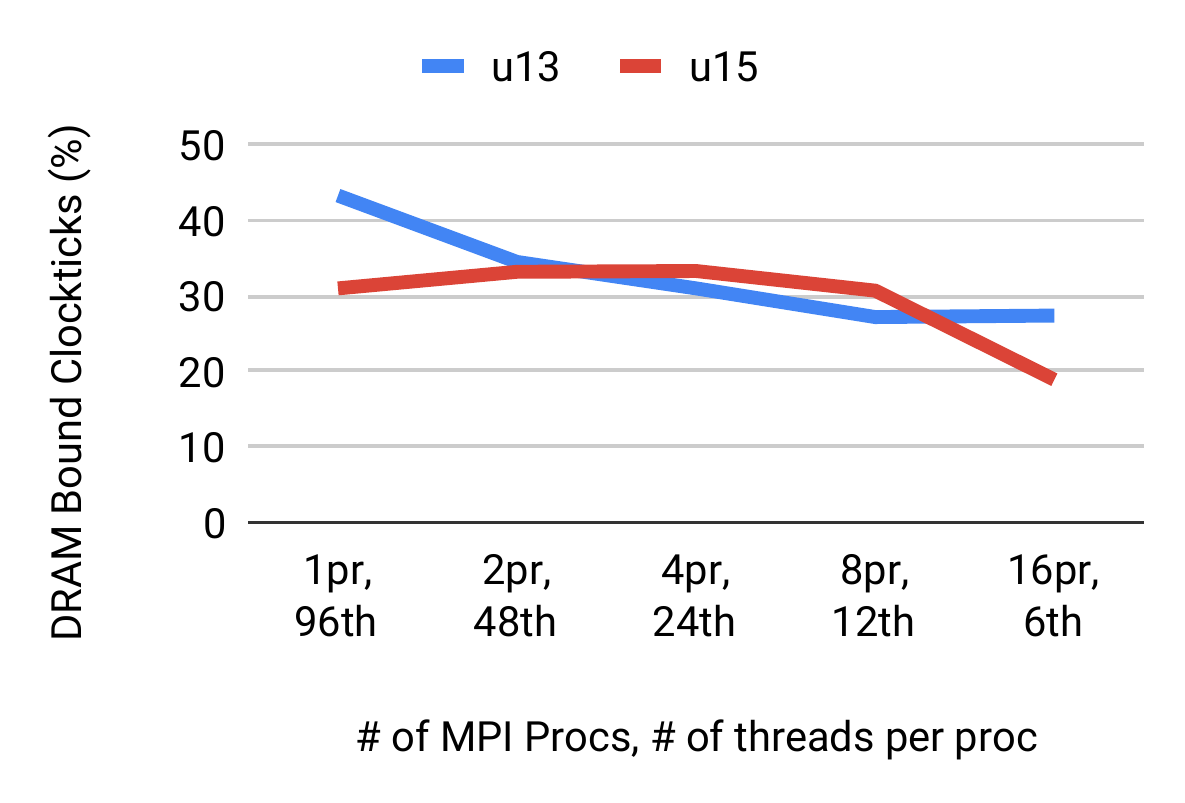}
\label{subfig:mpi-vtune-c}}
\subfloat[NUMA Remote Accesses]{\includegraphics[width=0.33\linewidth]{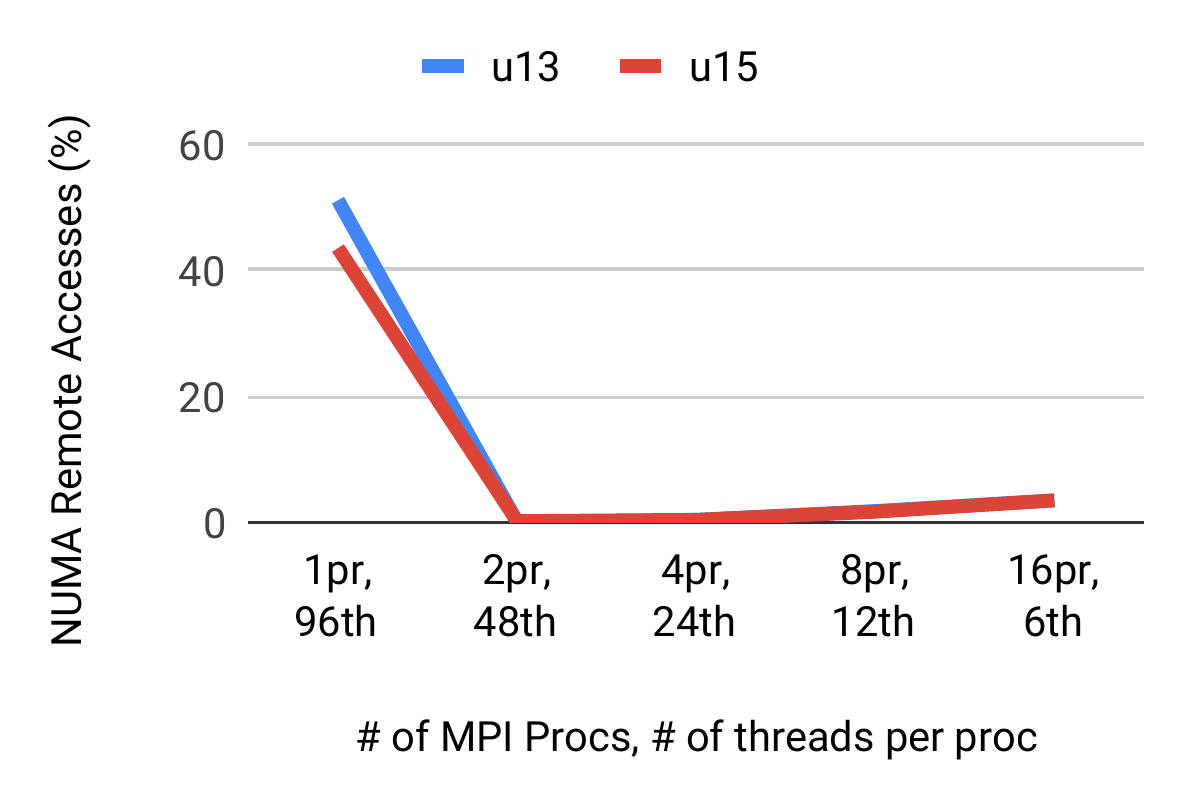}
\label{subfig:mpi-vtune-d}}
\caption{Performance metrics of Optane for distributed MPI version.}
\label{fig:mpi-vtune}
\end{figure*}

We measured hardware performance metrics of the machine for distributed MPI version, and the results are shown in Figure~\ref{fig:mpi-vtune}. The most important observation is that when going from single process to two MPI processes, percentages of NUMA remote accesses decreases sharply in Figure~\ref{subfig:mpi-vtune-d}. When counting 13 vertices subgraph, it goes down from 51.2\% to 0.1\%. When counting 15 vertices subgraph, it goes does from 43.6\% to 0.2\%. This implies that two MPI processes are placed into separate CPU sockets and they almost entirely use their own local DRAMs for data stores during the computation. This value remains very low even for larger number of MPI processes, since all MPI processes are placed in either one of two CPU sockets. Its maximum value is 3.5\% for 16 MPI processes. This seems to be the main reason for the better performance of the distributed version compared to the single version. Distributed version employs much better data locality in DRAM. 

Average DRAM bandwidth is lower when counting 15 vertices subgraph as shown in Figure~\ref{subfig:mpi-vtune-a} compared to the task for counting 13 vertices subgraph. The reason is that this task uses more memory than the DRAM capacity of the machine. Therefore, it uses persistent memory during computation for data stores. That reduces the average DRAM bandwidth since the persistent memory is slower and has less bandwidth compared to DRAM. On the other hand, 13 vertices case achieves much higher bandwidth since its data can fit entirely to the DRAM. This is also evident in Figure~\ref{subfig:mpi-vtune-b}. 13 vertices case operates with high DRAM bandwidth for higher percentages of execution times. 

Percentage of DRAM bound clockticks is very high for both tasks as shown in Figure~\ref{subfig:mpi-vtune-c}. It is around 30\% on the average. Although it decreases significantly for 13 vertices case when MPI processes are used, it increases slightly in the 15 vertices case for 2 and 4 processes. So, this metric does not seem to be strongly correlated by MPI usage.

In summary, it is very important to minimize NUMA remote accesses in Optane machine. In these tests, running multiple MPI processes eliminates NUMA remote accesses and the gains from that surpasses the costs of MPI inter-process communication and synchronization overheads. 

\subsection{Single Optane Node vs Conventional Cluster} \label{optaneVsCluster}

We compare the subgraph counting execution times on a single Optane node with a conventional cluster. The cluster nodes have 2x12-core Intel\textregistered\ Xeon\textregistered\ CPU E5-2670 v3 @ 2.30GHz processors and 128GB of memory. 

\begin{figure}
\centering
    
    \begin{minipage}{0.5\textwidth}
        \centering
        \includegraphics[width=0.95\textwidth]{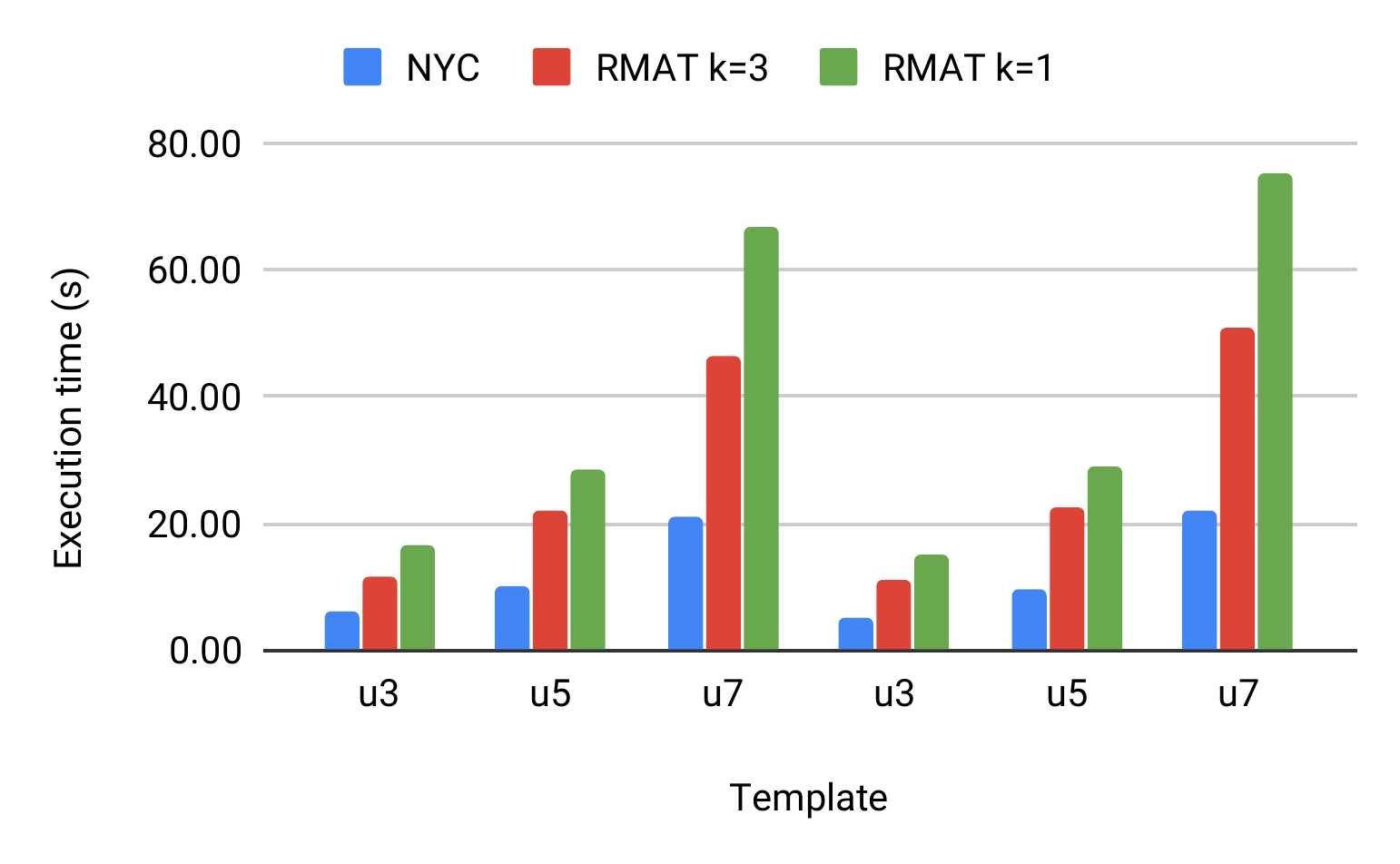}
    \end{minipage}
\caption{Comparing execution times on a single Optane node to a single cluster node}
\label{fig:singleNodeTests}
\end{figure}

First, we compare the running times on a single Optane node with a single cluster node. Since the cluster nodes have 24 CPU threads, we use 24 OpenMP threads on both machines. Figure~\ref{fig:singleNodeTests} shows the results. We run the algorithm on all three graphs with subgraphs that can fit into the memory of the cluster node with 128GB of memory. 

As Figure~\ref{fig:singleNodeTests} shows, both nodes provide comparable performance. Optane machine is around 10\% faster in larger subgraph counts. 

\begin{figure}
\centering
\includegraphics[width=0.5\textwidth]{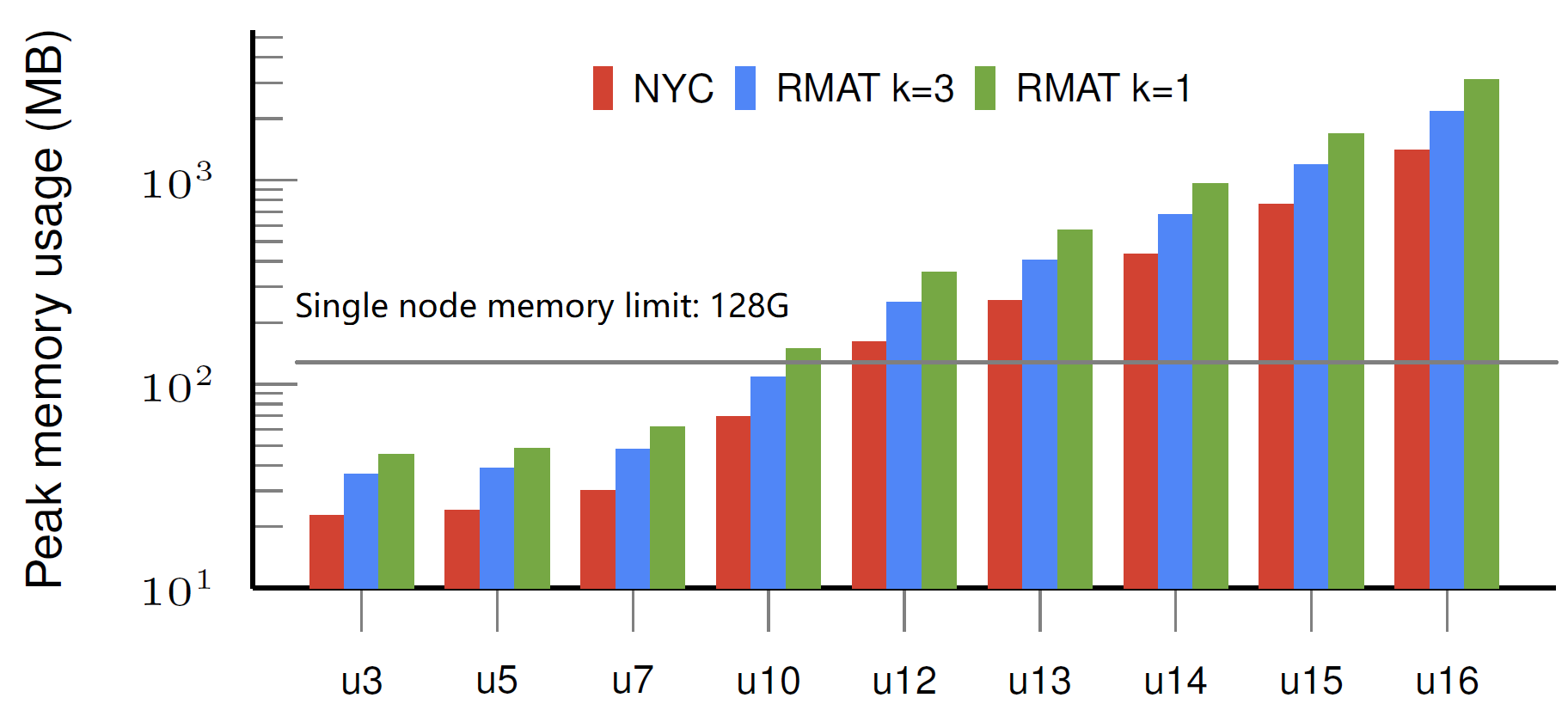}
\caption{Maximum memory usage comparison on different data sets. Data sets that exceed the memory limit will not run on a single node.}
\label{}
\end{figure}

Second, we compare the running times of the subgraph counting problem on a single Optane node with the distributed cluster nodes. We run the program in the cluster nodes by using the minimum number of nodes. The main limitation is the memory usage. Since each cluster node has 128GB of memory, we allocate a sufficient number of nodes to each run by determining the peak memory usage from single node Optane execution. For example, when we run the program with 12 vertices subgraph on nyc.graph, its peak memory usage is 158GB of memory. Therefore, we run it on two cluster nodes. In addition, we use the same number of threads in both environments. Since Optane has 96 threads, we use at most 96 threads in the cluster also. 

\begin{figure}
\centering
\subfloat[NYC]{\includegraphics[width=0.33\textwidth]{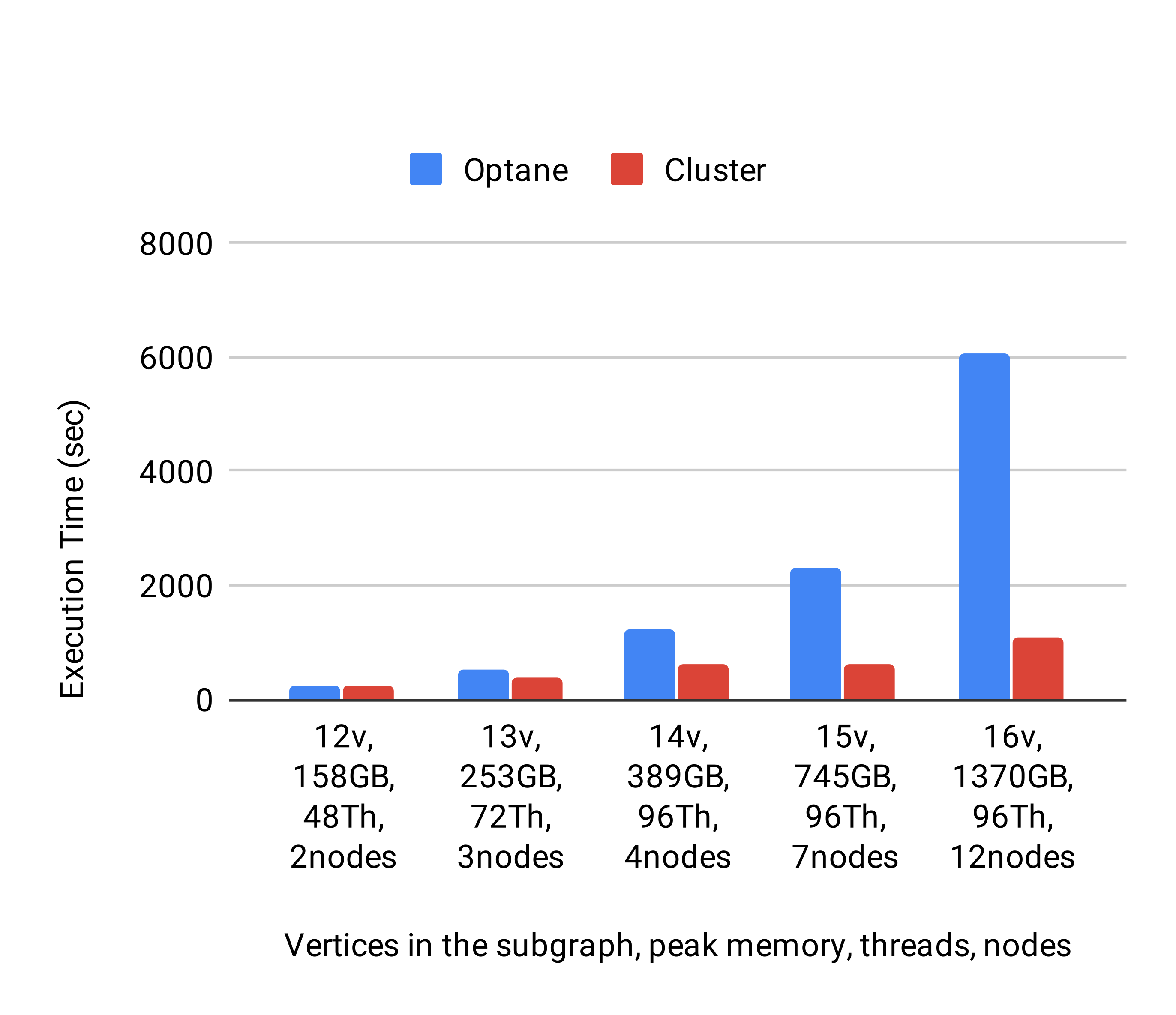} }
\subfloat[RMAT1]{\includegraphics[width=0.33\textwidth]{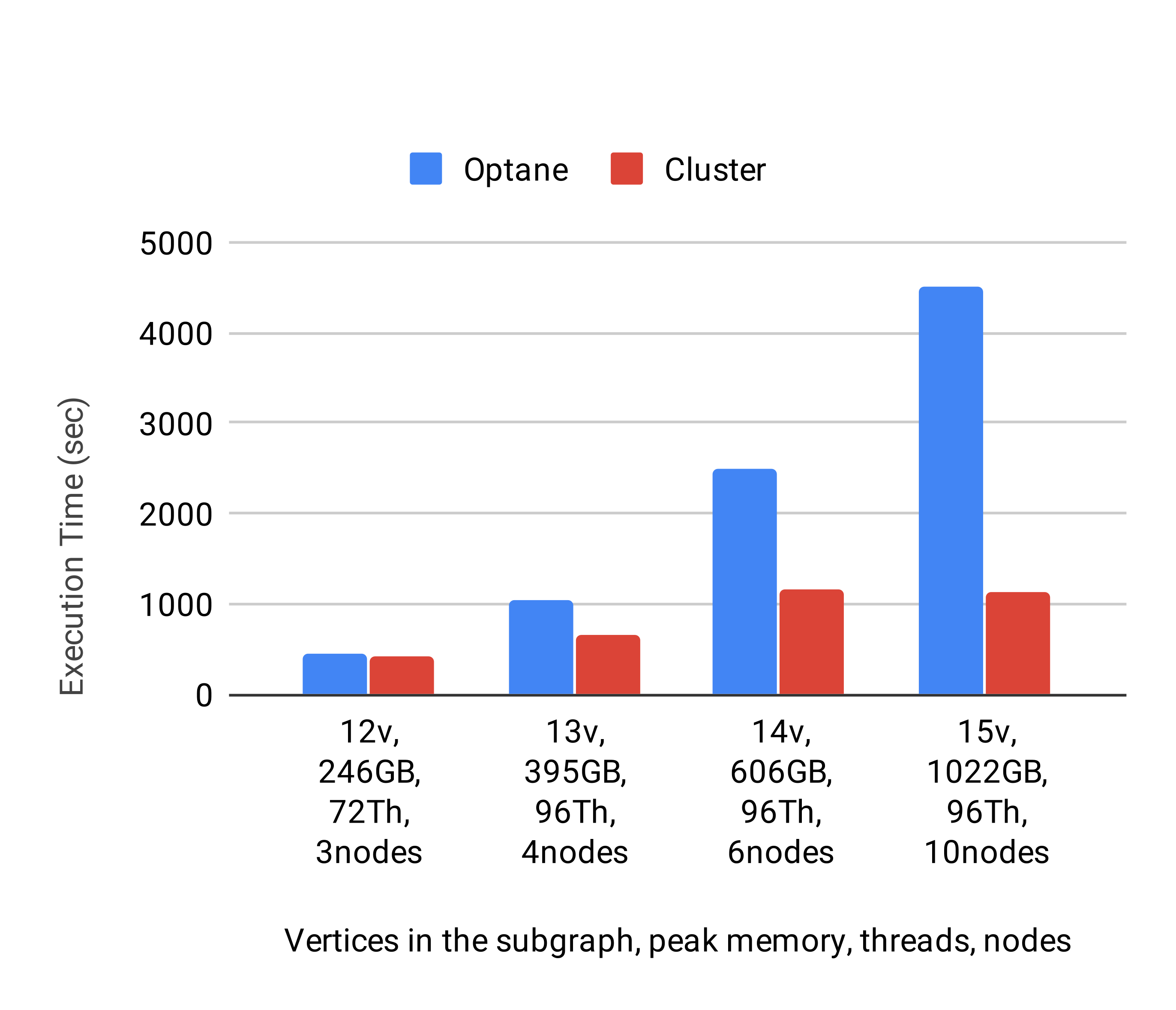} }
\subfloat[RMAT2]{\includegraphics[width=0.33\textwidth]{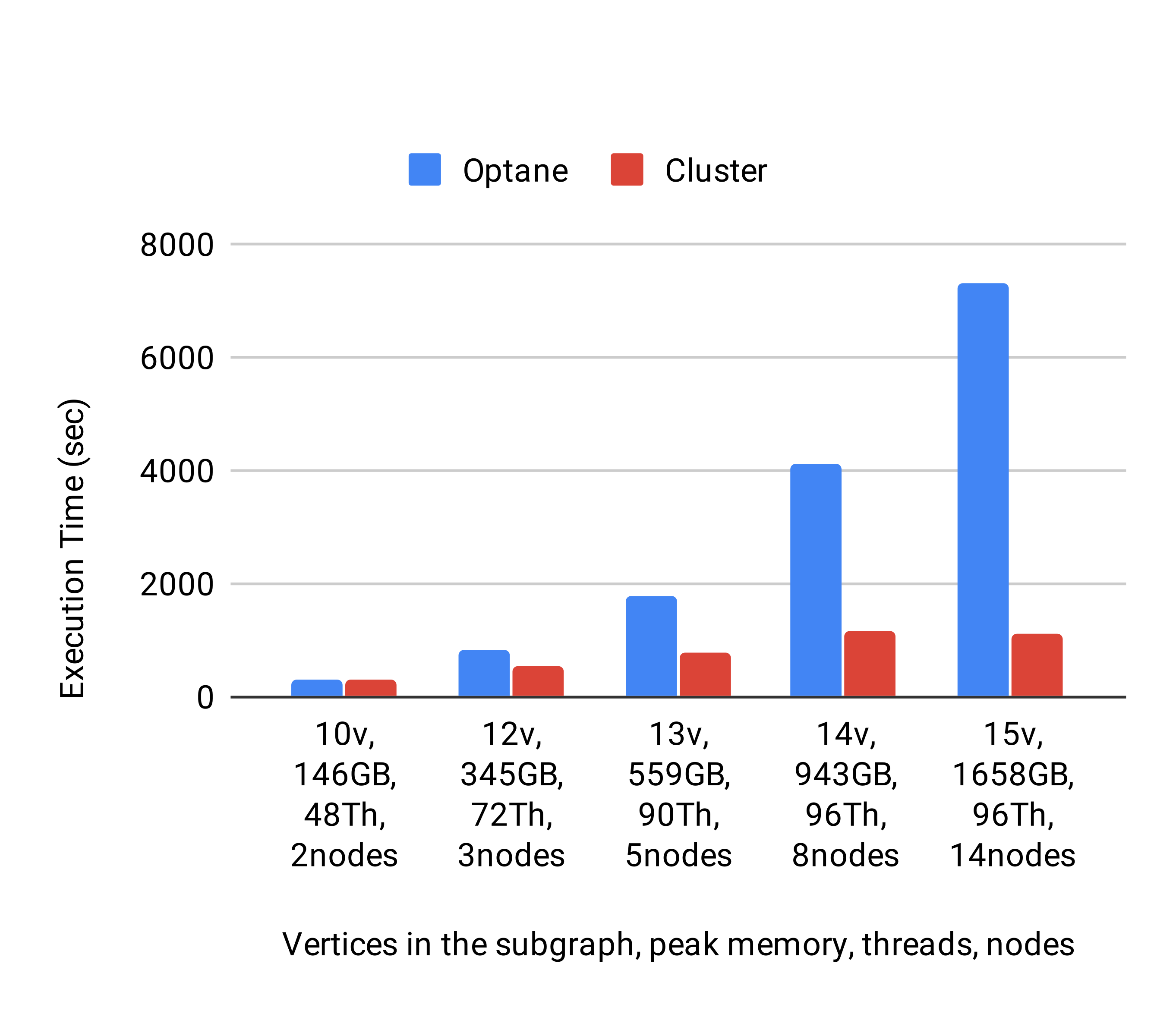} }
   
\caption{Comparing execution times on a single Optane node to many cluster nodes}
\label{fig:clusterNodeTests}
\end{figure}

Figure~\ref{fig:clusterNodeTests} shows the results. The label of each bar shows the program features: number of vertices in the subgraph template, peak memory usage, total number of threads used in both Optane and the cluster nodes, and the number of cluster nodes used. 

As the figures show, the distributed version that runs in the cluster with MPI provides much better performance. As the peak memory usage increases, the difference between cluster version and the Optane single node version also increases. For example, while counting 15 vertex subgraph on nyc.graph is 3.8 times faster in the cluster than the Optane one, counting 16 vertex subgraph in the cluster is 5.8 times faster. The same pattern exist in all three graph tests. For the largest subgraph tests, counting 15 vertices subgraph in RMAT1 graph is 4 times faster and counting it in RMAT2 graph is 7 times faster in the cluster. 

As the memory requirement increases, Optane machine is using persistent memory more and more. Its DRAM capacity is 384GB. When data can not entirely fit into this DRAM, it needs to move the data back and forth between the DRAM and the persistent memory. Since the persistent memory has both larger latency and smaller bandwidth, this results in slower performance for test cases that require larger memory footprints. On the contrary, in the distributed version, all data resides in DRAM of cluster nodes. Therefore, it is faster. The communication and synchronization overheads in the distributed version seem to be much less than the overhead of persistent memory in Optane machine. 

\section{Serverless Computing and Persistent Memory} \label{serverless}

Serverless computing platforms provide separate container images for each supported language. They use these images as the basis when constructing containers for functions. A separate container is created for processing each function. A function container goes through following life cycle stages. 
\begin{figure}
\centering
\includegraphics[width=0.5\linewidth]{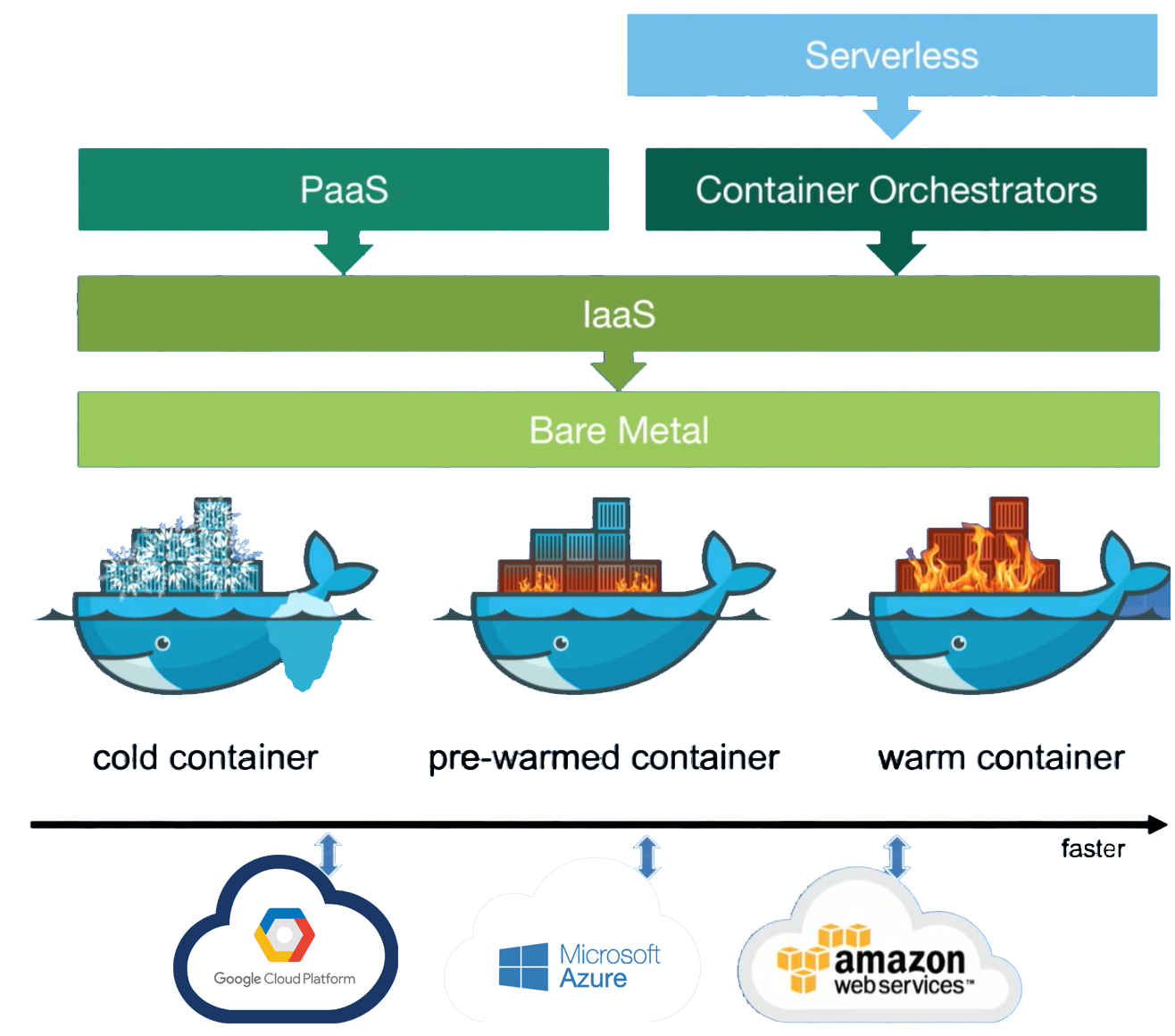}
\caption{The structure of serverless computing.\cite{serv}}
\end{figure}
\begin{itemize}
	\item \textbf{Cold starting the container}: A new container is created and initialized for the language of the function. For example, if the function is written in Java, a Java container is started. After this stage has completed, the container becomes \textit{prewarmed}. 
	\item \textbf{Warming up the container}: The function code is transferred to the container, dependencies are gathered, and any necessary initializations are performed. If the function code has not been compiled, it is compiled. The container is dedicated to this function from now on. The container becomes \textit{ready} or \textit{warmed}. 
	\item \textbf{Executing the function}: Input parameters are transferred to the container, the function code is executed and the result is transferred back to the user. 
	\item \textbf{Deleting the container}: If no function requests arrive for a predetermined amount of time, the container is deleted.
\end{itemize}

Serverless computing systems use caching methods to improve end-to-end execution delays \cite{owstart}. They have two levels of container caching. First, when a function is executed, its container is kept in memory in ready state for some period of time, before deleting it. It is assumed that subsequent requests are more likely for recently executed functions. Second, for some frequently used languages, a set of prewarmed containers are kept in memory for upcoming function requests. This avoids cold starting containers and reduces end-to-end execution delays. When a function request arrives, only warming up and function execution are performed. However, like in all caches, there are always cache misses and cold starting of a container is a serious issue in serverless computing systems \cite{colds}. 

\begin{table}[htb]
\centering
\begin{tabular} { | m{3.5cm} | m{6.5cm} | } 
 \hline
 Language and Version & Container Image Name \\ 
 \hline
 Node.js 10	& openwhisk/action-nodejs-v10:nightly \\ 
 \hline
 Ruby 2.5 & openwhisk/action-ruby-v2.5:nightly \\
 \hline
 PHP 7.3 & openwhisk/action-php-v7.3:nightly \\
 \hline
 Python 3.6 & openwhisk/python3action:nightly \\
 \hline
 Java 8 & openwhisk/java8action:nightly \\
 \hline
 DotNet 2.2 & openwhisk/action-dotnet-v2.2:nightly \\
 \hline
 Go 1.11 & openwhisk/actionloop-golang-v1.11:nightly \\
 \hline
 Swift 4.2 & openwhisk/action-swift-v4.2:nightly \\
 \hline
\end{tabular}
\caption{Used languages and container image names}
\label{table:containers}
\end{table}

We performed the tests using OpenWhisk serverless computing system. We used 8 languages listed at Table~\ref{table:containers}. The table also shows the container image names for each language. 

We performed the tests on this section on a Linux machine with 2x12-core Intel\textregistered\ Xeon\textregistered\ CPU E5-2670 v3 @ 2.30GHz processors and 128GB of memory. It has both an HDD and an SSD drive. We configured Docker to store all container images either in HDD or SSD drives when conducting the tests. Docker used overlay2 storage driver for both cases. We performed each test 5 times and averaged the results to minimize the effects of system performance variations. 

\subsection{Single Function Executions using HDD or SSD}

\begin{figure*}[htb]
\centering
\subfloat[Execution Delay on Ready]{\includegraphics[width=0.4\linewidth]{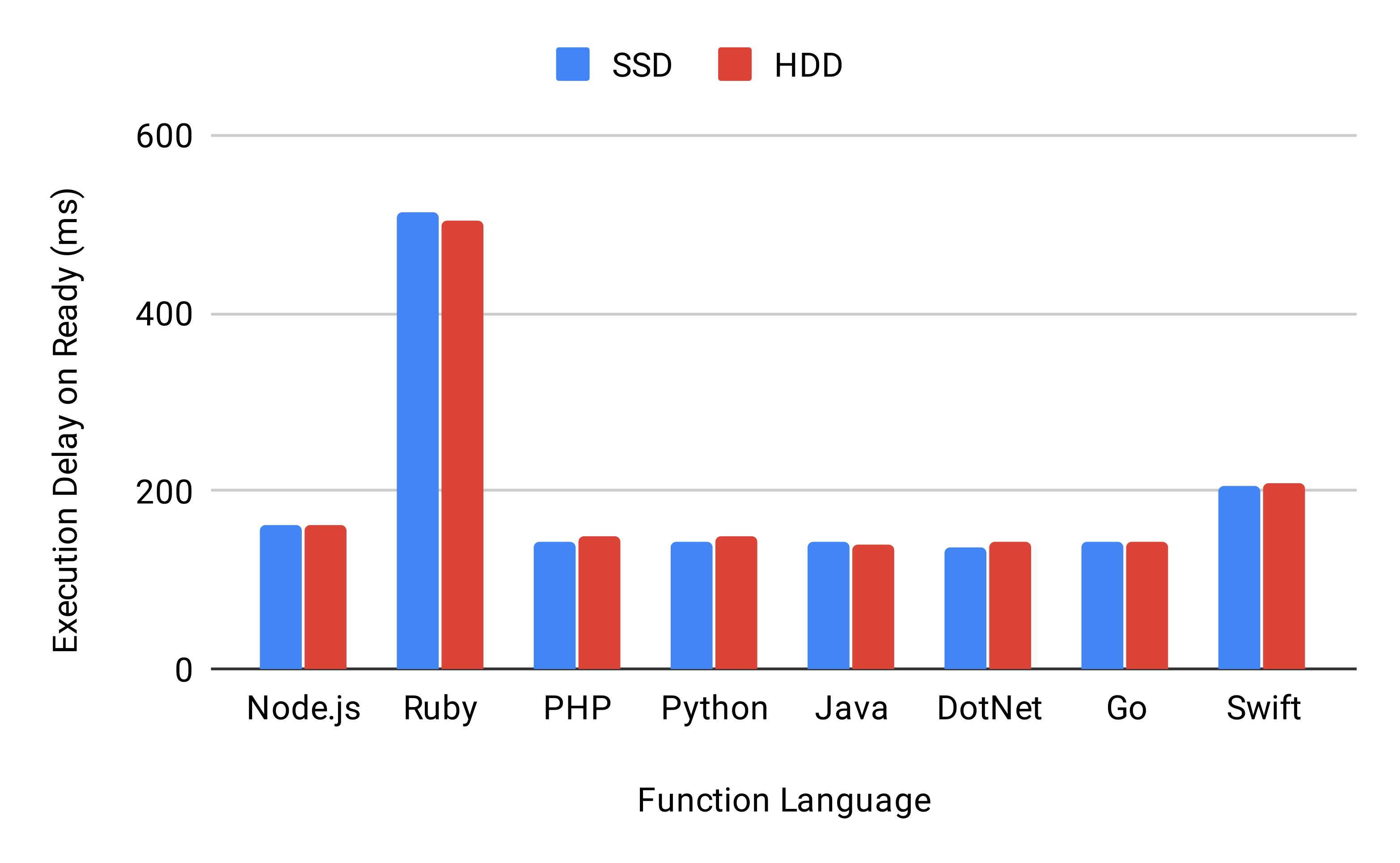}
\label{subfig:owhisk-fn-a}}
\hspace{5ex}
\subfloat[Prewarm to Ready Delay]{\includegraphics[width=0.4\linewidth]{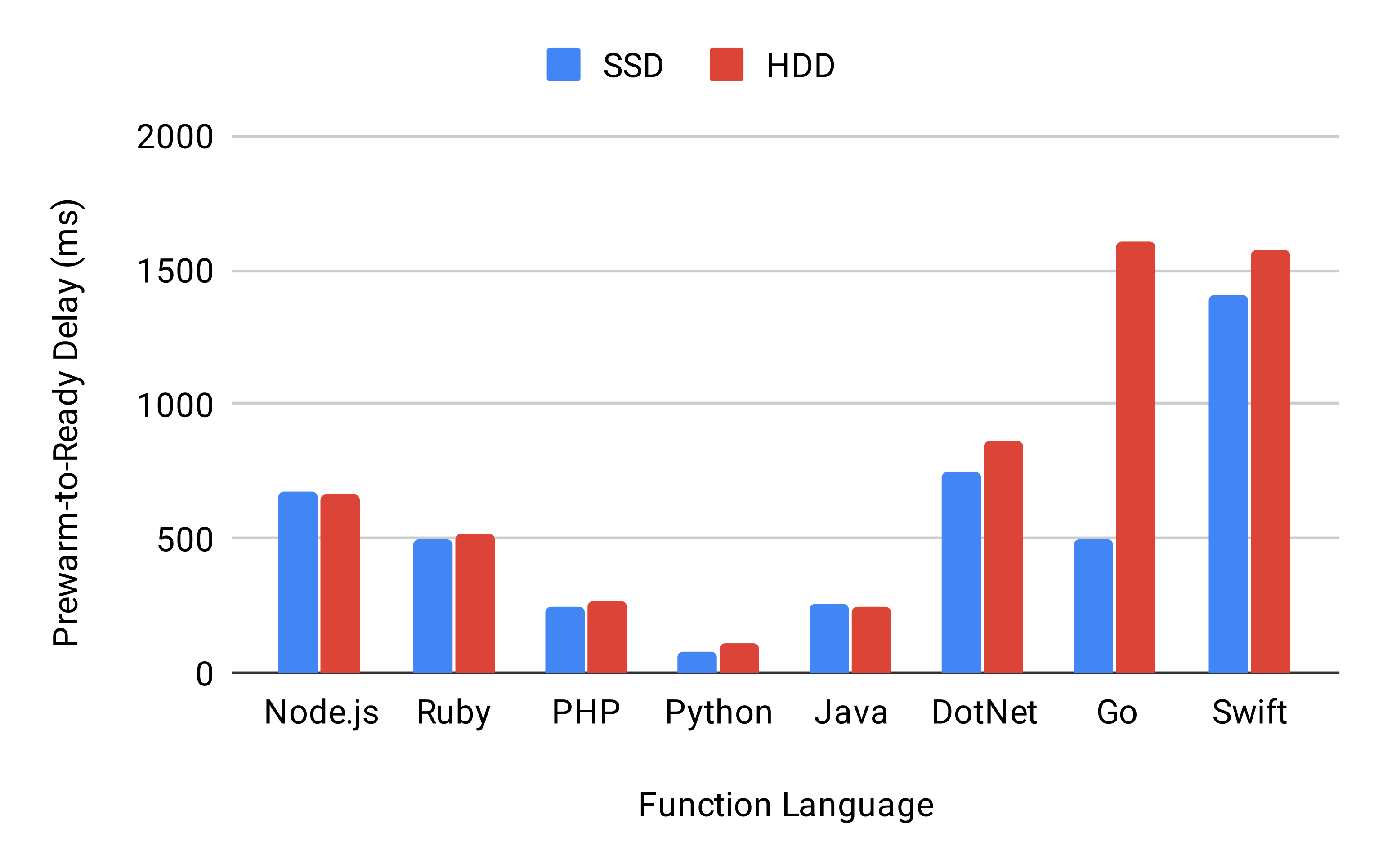}
\label{subfig:owhisk-fn-b}}
\vfill
\subfloat[Cold to Prewarm Delay]{\includegraphics[width=0.4\linewidth]{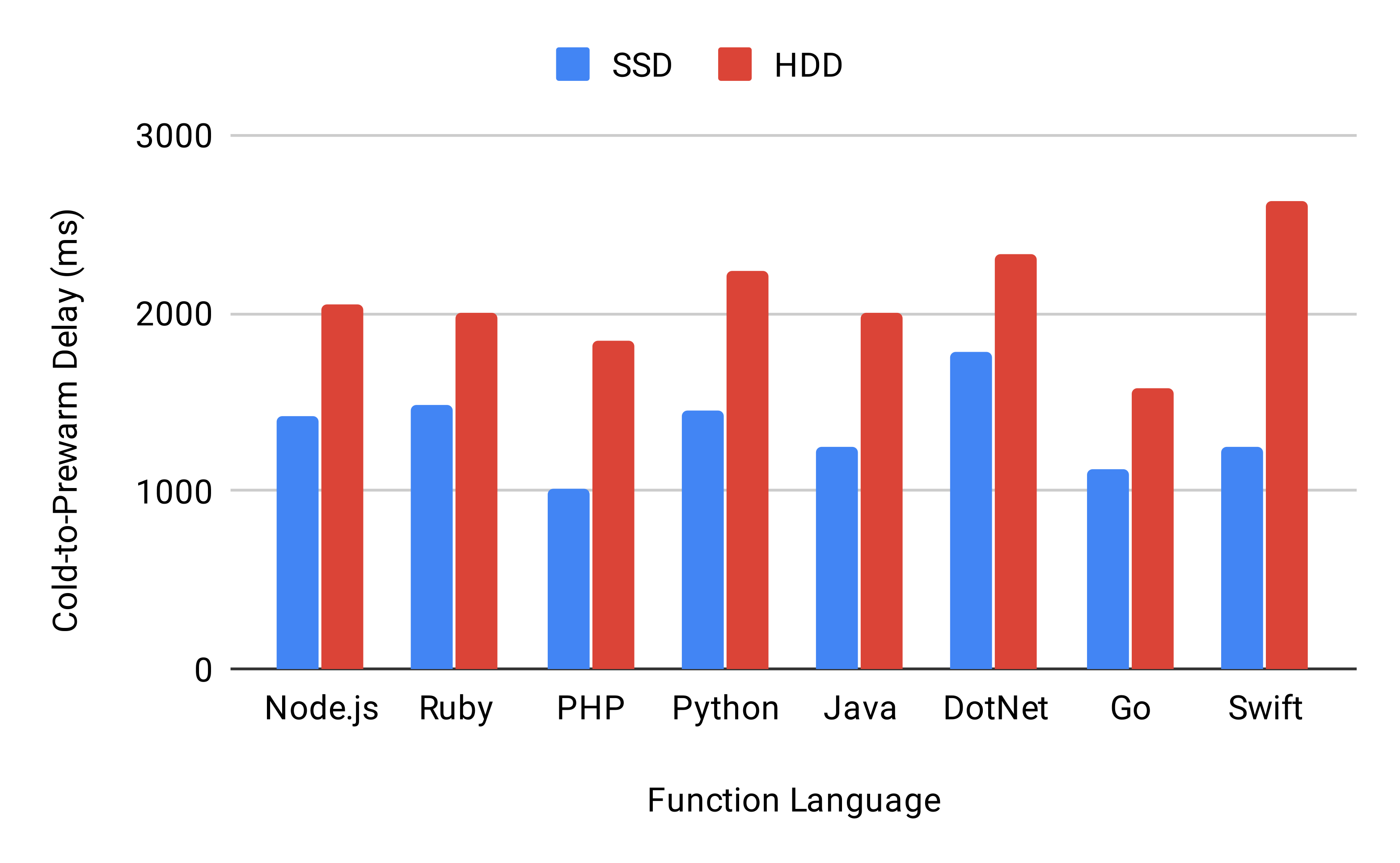}
\label{subfig:owhisk-fn-c}}
\hspace{5ex}
\subfloat[SSD Delay for stages]{\includegraphics[width=0.4\linewidth]{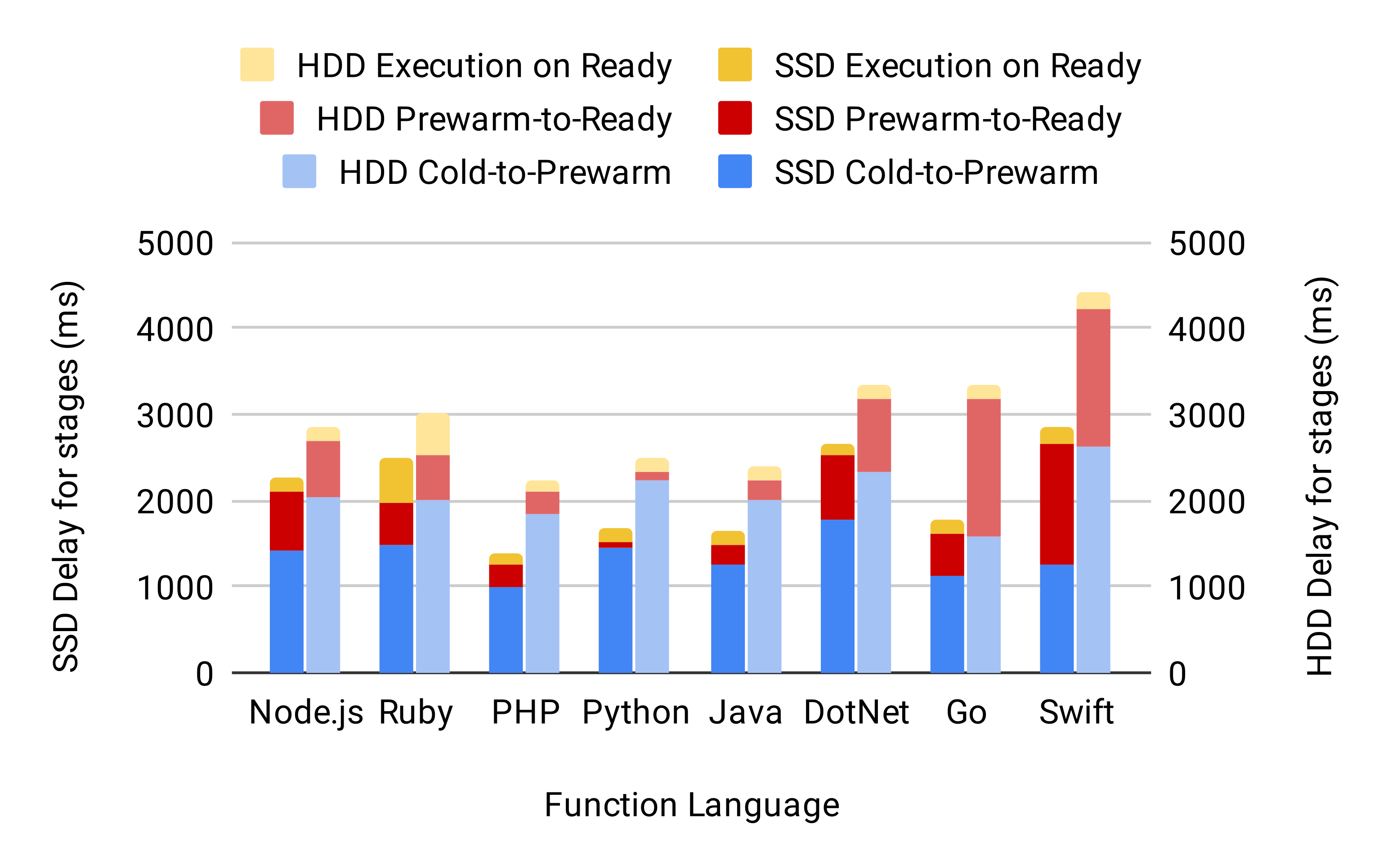}
\label{subfig:owhisk-fn-d}}

\caption{Stages of single function execution delays in OpenWhisk for 8 languages}
\label{fig:fn-delay}
\end{figure*}

We measured end-to-end single function execution delays using cold started, prewarmed and ready containers. The results are shown in Fig \ref{fig:fn-delay}. We used a simple Hello World function in 8 different languages to demonstrate the delays introduced by the system components. We sent a parameter to the function and received a response. The function code execution is not supposed to introduce any meaningful delays.

Figure~\ref{subfig:owhisk-fn-a} shows end-to-end function execution delays when there is a ready container for the requested function. This shows the delays introduced mostly by OpenWhisk system components. Average execution delay is 153ms with SSD and 156ms with HDD excluding the Ruby function. Surprisingly a simple function takes a few hundred milliseconds in Ruby. Function execution delays are almost the same for both storage types. As we expected, this implies that there is not enough storage activity during these function executions to make a difference in end-to-end delays.

Figure~\ref{subfig:owhisk-fn-b} shows prewarmed-to-ready delays; the time it takes to warm up containers. We measured end-to-end function execution delays on prewarmed containers and subtracted the end-to-end execution delays on ready containers. Prewarmed-to-ready delays show significant differences for these languages. While Python container is the fastest, Go and Swift are the slowest. There is no meaningful time differences for 5 languages based on the storage type. This implies that those containers do not perform much storage activity during this stage. For the remaining three containers (DotNet, Go, Swift), they perform significantly faster with SSD. The results imply that these 3 containers are performing a considerable amount of storage activity during this time. When we averaged the prewarmed-to-ready delays for all 8 languages, it is 551ms with SSD and 729ms with HDD. Therefore, depending on the container type, using faster storage may help readying prewarmed containers faster. 

Figure~\ref{subfig:owhisk-fn-c} shows cold-to-prewarmed delays. We measured end-to-end function execution delays on cold started containers and subtracted the end-to-end execution delays on prewarmed containers. This is the most significant delay introduced when executing serverless functions. It includes starting containers from scratch. Read-only container layers are loaded and the writeable container layer is created on top of the other layers. Therefore, this is the stage with most storage activity. On the average, it takes 1346ms to prewarm a container with SSD and 2084ms with HDD. Therefore, on the average, cold starting containers are 35\% faster when SSD storage is used. 

Figure~\ref{subfig:owhisk-fn-d} shows end-to-end function execution delays with stages, combining all these three delays. Left bars of each pair show the delays with SSD and right bars show the delays with HDD. Average function execution delay for 8 functions is 2095ms with SSD and 3012ms with HDD for cold starts. Functions are on the average executed 30\% faster on cold containers when SSD drives are used. This is a very significant improvement for serverless computing. It would improve the responsiveness of the system considerably. Using persistent memory would increase the end-to-end execution delays even further, since they provide much faster storage access rates with higher bandwidth. 

\subsection{Concurrent Function Executions using HDD or SSD}

We test the performance of OpenWhisk for executing many functions at once. We would like to see how much difference does it make to use a faster storage system in case of bursty traffic. While the system is initially free without any running containers, we send many function execution requests in parallel. Each function is different and only one request is submitted for each function. Therefore, each function initiates a cold start of its container. We measure the total execution times for these functions using SSD or HDD drives. 

\begin{figure}
\centering
\includegraphics[width=0.5\textwidth]{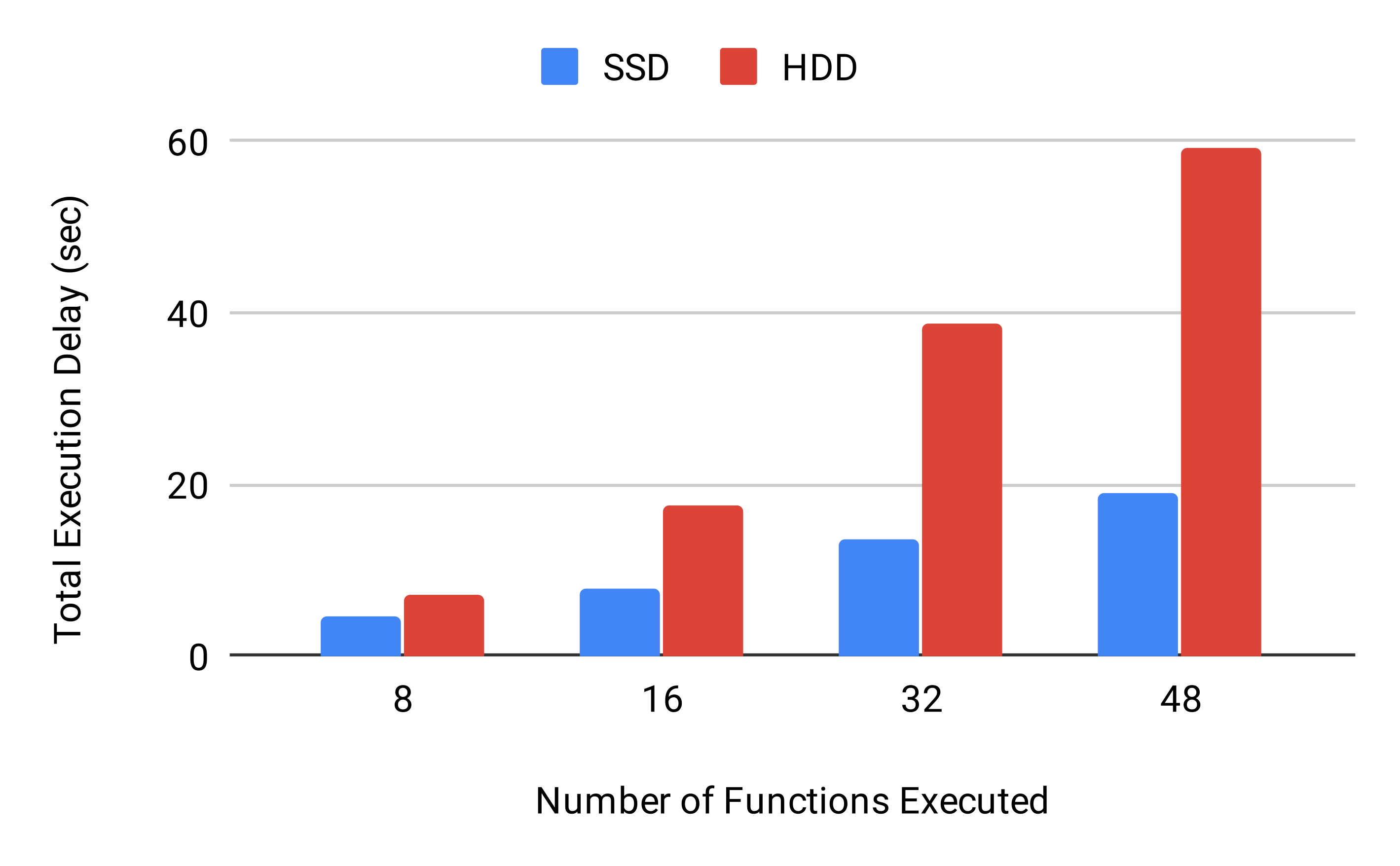} 
\caption{Concurrent function executions with cold starts using SSD or HDD drives in OpenWhisk}
\label{fig:owhisk-many}
\end{figure}

The results are shown in Fig \ref{fig:owhisk-many}. First pair of bars show the total time spent executing 8 functions; one function for each language. Second pair of bars show the total delay for executing 16 functions; two functions for each language, etc. As the results show, serverless system performs much better with SSD storage. As the number of concurrent requests increase, the advantage of SSD grows further. When 8 concurrent requests processed, it takes 4.5 seconds with SSD and 7.3 seconds with HDD. Therefore, processing 8 functions with HDD takes almost two times longer. However, when 48 concurrent functions are processed, it takes 19.1 seconds with SSD and 59.2 seconds with HDD. Therefore, processing 48 requests with HDD takes more than 3 times longer. 

These results show that higher bandwidth SSD storage improves serverless system responsiveness significantly in the case of heavy traffic with cold starts. Running serverless systems on persistent memory machines may improve further the quality of concurrent function executions with their higher bandwidth capacity. 

\subsection{Related Work}

Performance of serverless computing systems have been investigated by some recent papers. Lloyd et al. investigated serverless function performances on AWS Lambda and Azure Functions ~\cite{colds}. They examined response times with cold and warm containers, the impact of load balancing, container caching policies, and memory reservation sizes on performance. Lee et al. investigated the performance of concurrent execution of functions in AWS Lambda, Azure Functions, Google Functions and IBM OpenWhisk ~\cite{par-sc}. They examined concurrent function throughput, concurrency of CPU intensive, disk I/O intensive, and network I/O intensive functions. Klimovic et al. designed an elastic, distributed data store for serverless computing systems ~\cite{pocket}. They leveraged multiple storage technologies such as DRAM, SSD, and HDD to minimize the cost while ensuring optimized I/O performance. They tested the designed system on AWS cloud using EC2 instances. 

Intel\textregistered\ Optane\textsuperscript{TM} DC Persistent Memory Module is a new storage technology with great promise to improve many aspects of applications in servers. It provides larger memory capacities and much faster persistent storage. There have been many recent studies investigating various aspects of this new technology. Izraelevitz et al. performed an in-dept performance review of this module with its capabilities as main memory and a persistent storage device \cite{optanePerf}. They compared its latency and bandwidth characteristics to DRAM values. Gill et all. investigated the performance of a single Optane machine for massive graphs analytics  \cite{pm-graph}. They compared the performance of a single Optane machine with a distributed set of cluster nodes for graph analytics. Mahapatra et al. designed efficient data structures such as Doubly Linked List, B+Tree and Hashmap for persistent memory machines by removing redundant fields \cite{pm-ds}. Van Renen et al. conducted performance evaluations of persistent memory in terms of bandwidth and latency, and developed guidelines for efficient persistent memory usage for two essential I/O primitives in disk-based database systems: log writing and block flushing \cite{pm-db}.

Our planned work is distinct from above-mentioned studies. We plan to investigate the performance improvements in serverless computing systems by using faster and higher bandwidth persistent storage. 

\section{Conclusions} \label{conclusions}

Our results for subgraph counting show that a single Optane machine can conveniently run a multi-threaded memory-intensive application with terabytes of memory usage. However, to get the best performance out of Optane machine, multi-threaded programs need to be NUMA aware. The impact of NUMA remote accesses is significant on execution times. The results in section~\ref{scalingWithMPI} show that the distributed version of the subgraph counting algorithm on Optane may run 34\% faster, despite the fact that it additionally introduces some inter-process communication and synchronization overheads. The main reason for this is the fact that the distributed version has much lower NUMA remote access rates.  

Our results for serverless computing tests show that using SSD drives improves the end-to-end execution delays of single functions with cold starts significantly. While on the average, executing a simple function takes 2.1 seconds with SSD, the same function takes 3 seconds with HDD. Therefore, using SSD improves the responsiveness of the serverless computing system considerably. In addition, when SSD drives are used, concurrent functions are executed much faster. When 48 concurrent function requests are submitted, they are processed 3 times faster with SSD. Therefore, using faster storage systems significantly improve the end-to-end delays of single function executions and provide much better handling of heavy function loads. These tests demonstrate that using Optane persistent memory machine for serverless computing will provide even faster end-to-end function executions and much faster processing of bursty function requests. 

\subsection{Future Work}
We are planning to test various OpenMP thread binding methods to reduce NUMA remote accesses in the single process version. Further we will look into the performance metrics of the memory hierarchy in more detail and perform roof-line modeling analysis to see whether we can reduce the percentages of DRAM bound clockticks. We are also planning to test with other memory-intensive problems with different distributed communication and load sharing features. Our overall objective is to determine the most important factors when running memory-intensive applications on a single Optane machine. 

To serverless computing tests on Optane machine using persistent memory in app-direct mode, We would like to investigate the impact of faster and higher bandwidth persistent memory on these systems. We plan to investigate the performance improvements when cold starting and concurrently initializing Docker containers; executing simple functions, fat functions with larger memory footprints, I/O intensive functions, many concurrent functions, and sequence of functions in serverless systems. We want to compare their performance on HDD, SSD, and persistent memory.

\end{document}